# On the variability of the reflection nebula Van den Bergh 27 surrounding RY Tau

Harald Strauß[1], Klaus Bernhard[2]

1) Astronomischer Arbeitskreis Salzkammergut
email: h.strauss@aon.at
2) Bundesdeutsche Arbeitsgemeinschaft für Veränderliche Sterne e.V., Germany
American Association of Variable Star Observers (AAVSO), USA
email: klaus.bernhard@liwest.at , ORCID 0000-0002-0568-0020

**Abstract:** We present an analysis of the short- and long-term optical variability in Van den Bergh 27 (vdB 27), the reflection nebula surrounding RY Tau. The observed variations form a light-echo pattern, with apparent propagation speeds varying across different regions of the nebula and reaching up to approximately 3.6 c, consistent with geometric projection effects. The observed variations in nebular surface brightness are consistent with changes in illumination from the central star.

## 1 Introduction

Reflection nebulae are interstellar clouds that shine by scattering the light of nearby stars rather than by producing their own radiation. Since shorter wavelengths are scattered more efficiently, these nebulae often show a bluish appearance, as illustrated by well-known objects such as the Pleiades. In contrast to emission nebulae, the central stars do not ionize the surrounding gas but instead illuminate dust particles in their vicinity [1].

For a long time, reflection nebulae were often assumed to be optically constant, largely owing to the lack of systematic time-domain observations. Only subsequently was it recognised that a small subset exhibit pronounced changes in brightness and apparent morphology on timescales of weeks to months. Classic examples of these so-called variable reflection nebulae include Hubble's and Hind's variable nebulae [2, 3]. For these objects, the observed variability is commonly attributed either to intrinsic variations of the illuminating source or to time-dependent shadowing effects caused by circumstellar dust structures.

In 2025, we detected rapid brightness variations on timescales of a few days in the reflection nebula vdB 24 ("StraBe 1") surrounding XY Per, adding it to the small group of known variable reflection nebulae [4]. Extending this approach to other young, strongly variable stars, we subsequently identified optical variability in the nebulae around V1818 Ori ("StraBe 2") and NSV 16694 ("StraBe 3"), with characteristic timescales of weeks to months [5].Together, these findings suggest that time-dependent illumination and shadowing effects in circumstellar environments may be more common among reflection nebulae than previously assumed.

In this paper, we analyse the optical variability in vdB 27 identified using the same observational approach as applied to our previously studied objects. While this variability initially became apparent through our analysis of multi-epoch imaging data, a subsequent review of the literature revealed that vdB 27 has already been recognised as a variable reflection nebula [6]. Nevertheless, the object has remained comparatively little studied in a time-resolved context, particularly on the timescales addressed here.

## 2 Observation and data analysis

As part of our systematic survey, we investigated the variability of the reflection nebula vdB 27 using the procedure described in BAVJ 104 [4]. As an initial step, candidate reflection nebulae were selected from the CCD Guide catalogue compiled by the Astronomical Working Group Salzkammergut[1]. Archival images from the Digitized Sky Surveys[2] (DSS1 and DSS2) were then retrieved and directly compared, allowing us to identify changes in brightness and apparent morphology in the immediate vicinity of the illuminating star. This comparison provided clear evidence for optical variability in vdB 27. An animated sequence of the DSS r-band images is available at Zenodo: https://zenodo.org/records/18359790

For a more detailed analysis of the brightness behaviour of vdB 27, we examined imaging data obtained with the Zwicky Transient Facility, which has been conducting a systematic time-domain survey at Palomar Observatory since 2017. The ZTF camera, featuring e2v CCD231-C6 devices, is mounted on the Samuel Oschin Schmidt Telescope. Covering three passbands (g, r, and partially i), it reaches a limiting magnitude of about 20.5 mag, making ZTF data highly suitable for investigations of variable objects [7–9]. Unfortunately, the central star is already overexposed in the ZTF images. Therefore, other surveys, namely KWS (Kamogata/Kiso/Kyoto Wide-field Survey) and ASAS-SN (All Sky Automated Survey for SuperNovae) were used to compare the brightness variation of the nebula with that of the central star [10-12]. As a result, the photometric data do not correspond exactly to the dates of the ZTF images. Nevertheless, they are adequate for comparing the overall brightness evolution of the central star with that of the nebula.

## 3 Results

According to the entry in the International Variable Star Index (VSX) database, RY Tau is a classical T Tauri star of spectral type F8Ve-K1IV/Ve(T). It exhibits strong photometric variability, with its brightness ranging from approximately 9.3 to 11.7 V (range from AAVSO data) and several possible (semiregular) periods ranging from days (5.6 d) to several years. These variations are typical of T Tauri stars and are generally attributed to a combination of variable accretion from the circumstellar disk, hot and cool surface spots linked to strong magnetic activity, and changing extinction caused by inhomogeneities in the surrounding dust. However, RY Tau exhibits an unusually large variability amplitude for this class. A comprehensive discussion of RY Tau and its surrounding disk can be found in Ref. [13]. Two overview images with different levels of detail are presented in Figures 1 and 2. Figure 1 (Thomas Henne) illustrates the broader surroundings of the region, including the associated dark clouds. At the center of the large molecular cloud complex located in the constellation Taurus lies the prominent dark cloud Barnard 7, within which the significantly darker region Barnard 10 is embedded. To the east lies the nebular complex Barnard 214, whose southern end hosts the reflection nebula vdB 27, illuminated by the classical T Tauri star RY Tauri. Figure 2, based on archival Digitized Sky Survey (DSS) data obtained via the Aladin Sky Atlas, shows the immediate vicinity of the object.

[1] https://ccdguide.com
[2] https://skyserver.sdss.org/dr5/en/proj/advanced/skysurveys/poss.asp

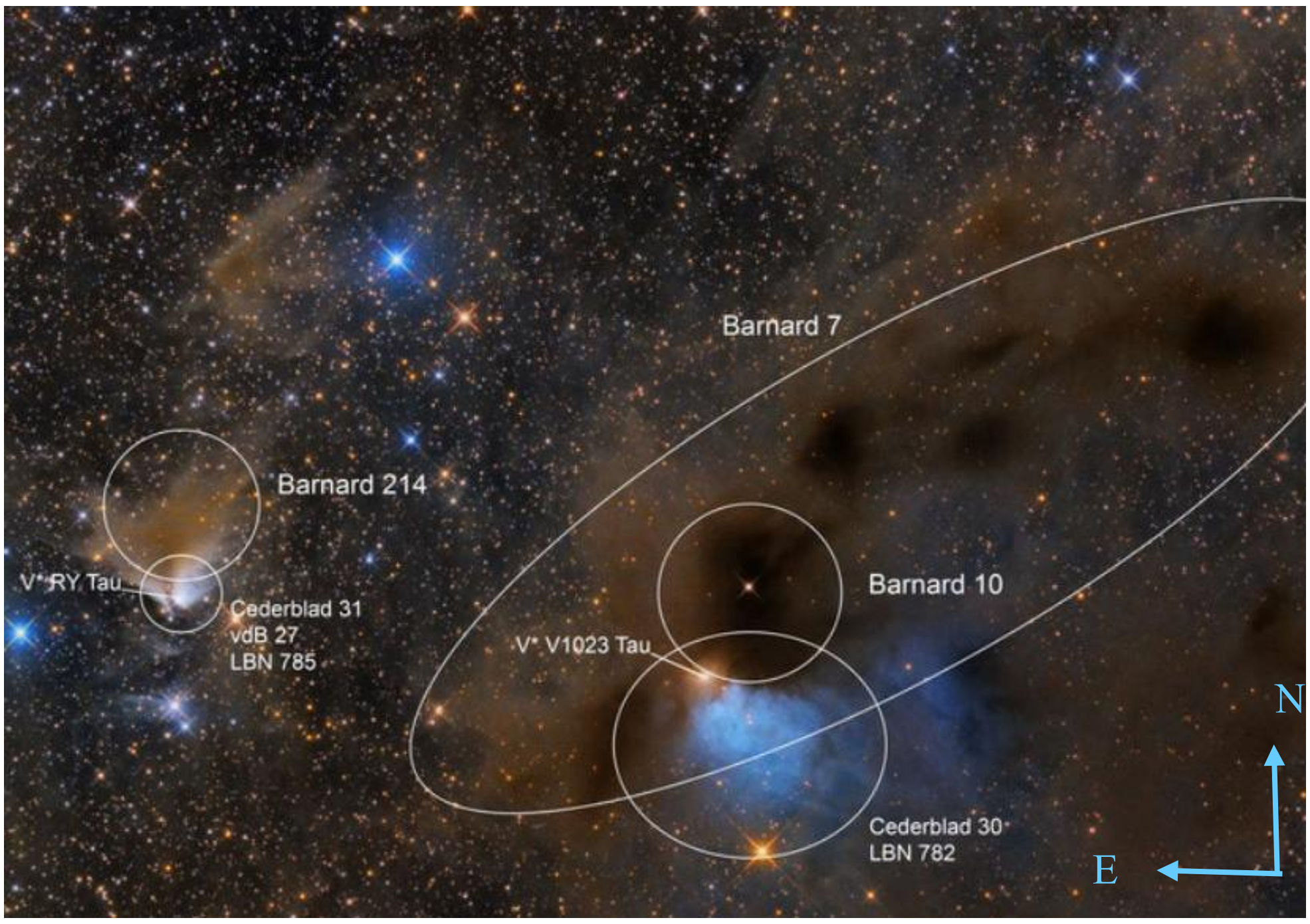

Figure 1: Overview of the field, RY Tau and its surrounding vdB 27 are situated close to the left edge of the image. - Thomas Henne | CCDGuide.com

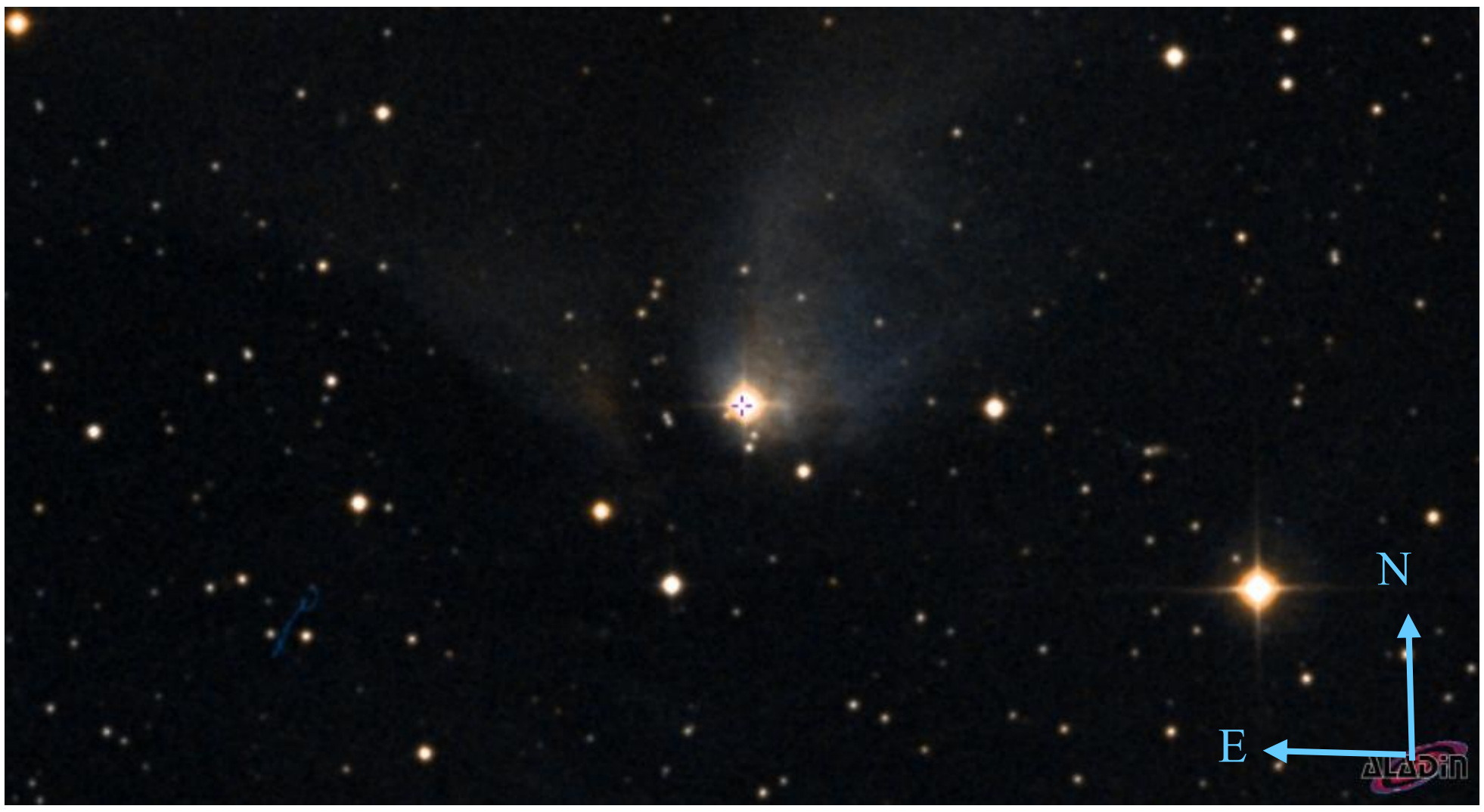

Figure 2: Overview DSS colour image of vdB 27, 13x7 arcmin

The essential data of RY Tau are listed in Table 1:

Table 1: Essential data of RY Tau

| | | |
|---|---|---|
| AAVSO VSX [14]: | Variability type | CTTS/ROT (T Tauri star) |
| | Identifiers | ASASSN-V J042157.38+282634.6, GSC 01828-00129 |
| | Magnitude Range | 9.3 - 11.7 V (range from AAVSO data) |
| Gaia DR3 [15]: | Gaia DR3 164551162164119424 | |
| | Right Ascension (J2000) | 04 21 57.4092* |
| | Declination (J2000) | +28 26 35.5553* |
| | Plx | 7.2349±0.2031 mas |
| | Gmag | 10.705 mag |

* All coordinates are taken from the Gaia DR3 catalogue (http://vizier.u-strasbg.fr/viz-bin/VizieR?-source=I/355). The coordinates (epoch J2000) are computed by VizieR, and are not part of the original data from Gaia (note that the coordinates are computed from the positions and the proper motions).

**Variability of the central star**

In Figures 3 and 4, the light curves from the KWS survey and from ASAS-SN are shown.

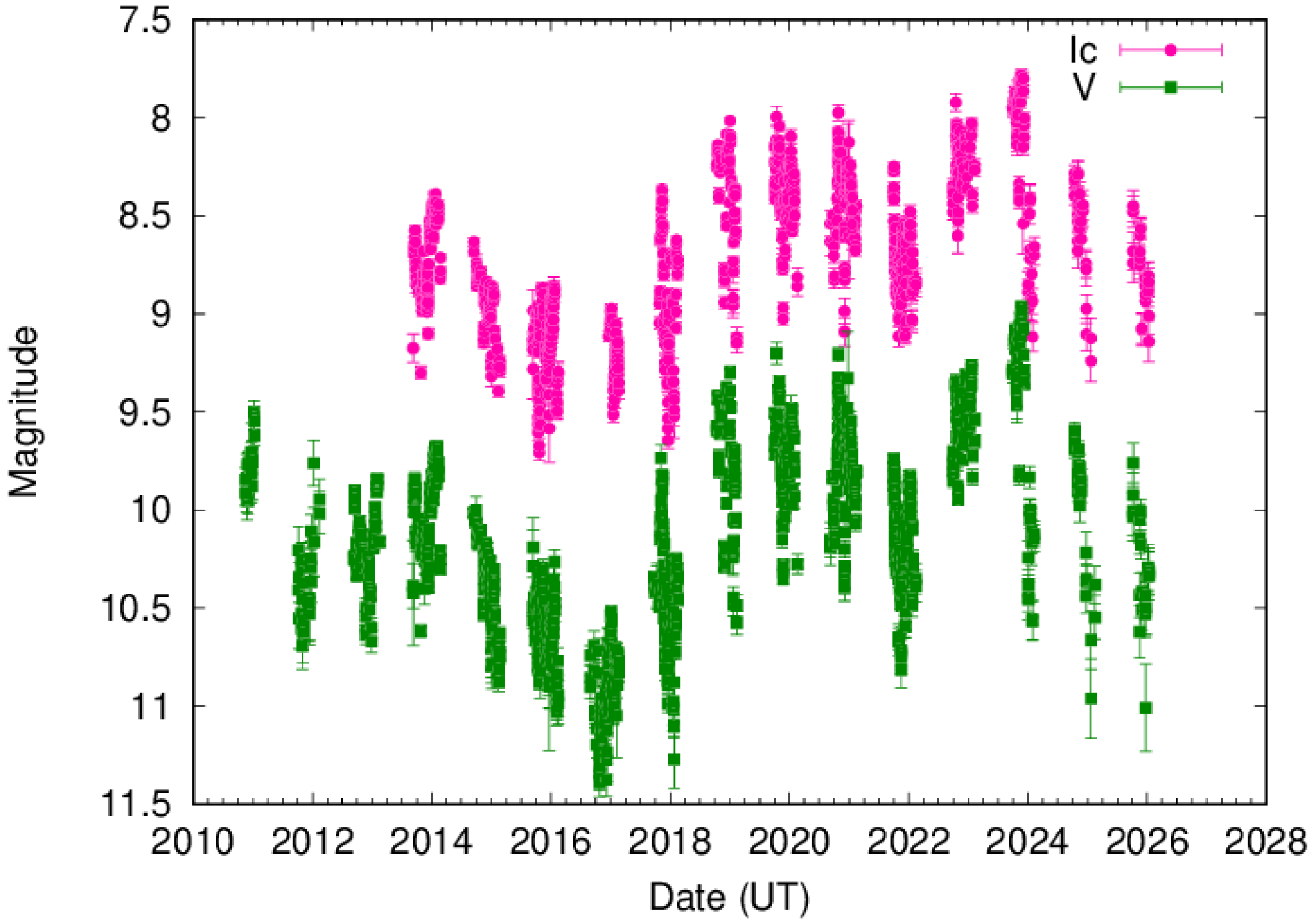

Figure 3: KWS Ic and V light curve of RY Tau between 2011 and 2025

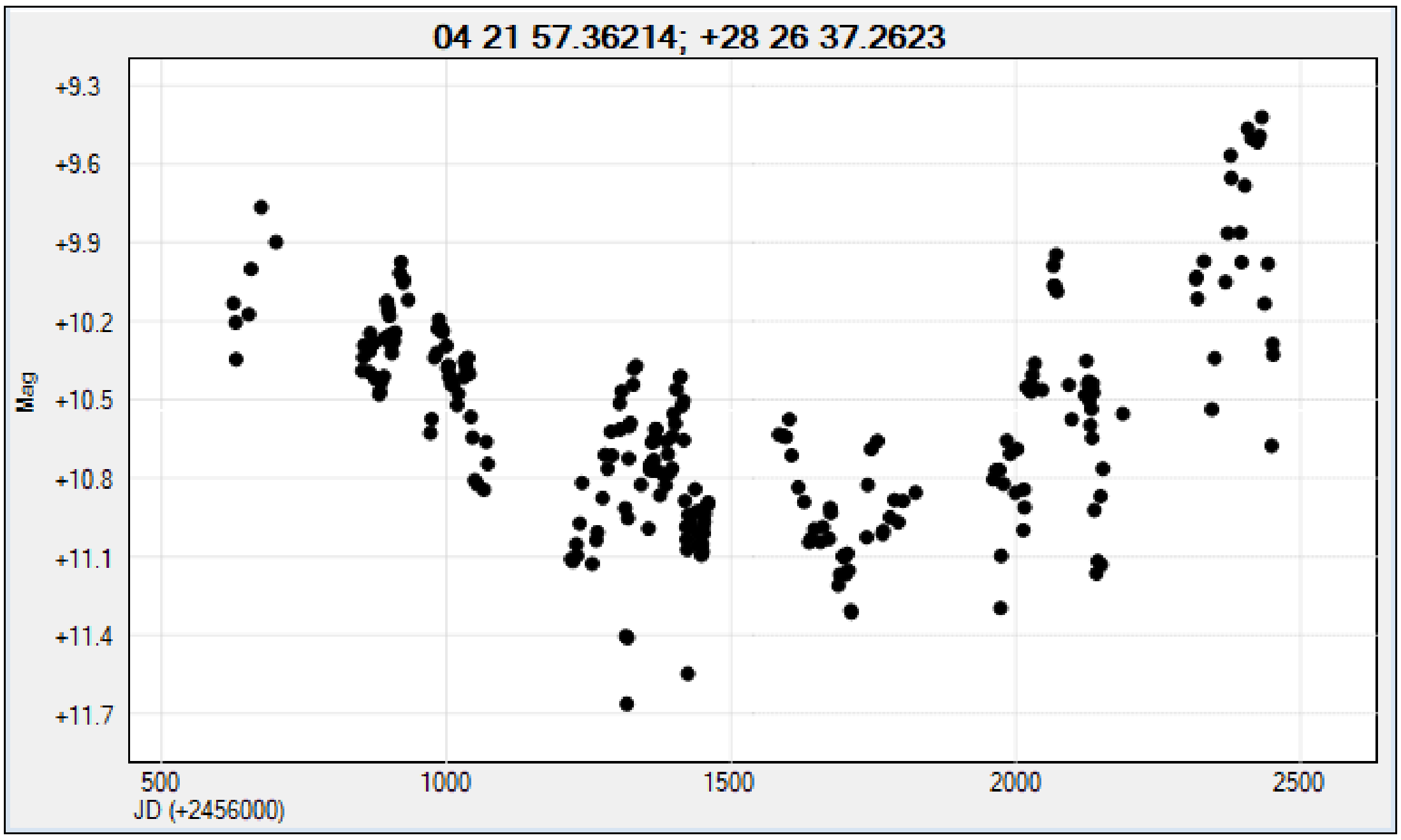

Figure 4: ASAS-SN V light curve of RY Tau between 2012 and 2019

The light curve consists of a long-term component with characteristic timescales of years and a shorter component on the order of weeks to months. A period analysis of the KWS-V and of the ASAS-SN V data was performed using the Lomb–Scargle GLS method in Peranso [16]. Prior to the analysis, only obvious outliers were removed, while no detrending was applied. In the GLS periodograms, the test statistic (Theta) represents the normalized power of the signal as implemented in Peranso. The periodograms shown in Figures 5 and 6 reveal several broad peaks in the KWS data in the range of approximately one year (maximum Theta at ~400 days), which are not present in the ASAS-SN data. This indicates that the signal may represent only a transient periodicity, or that it is caused by the sampling and a one-year alias.

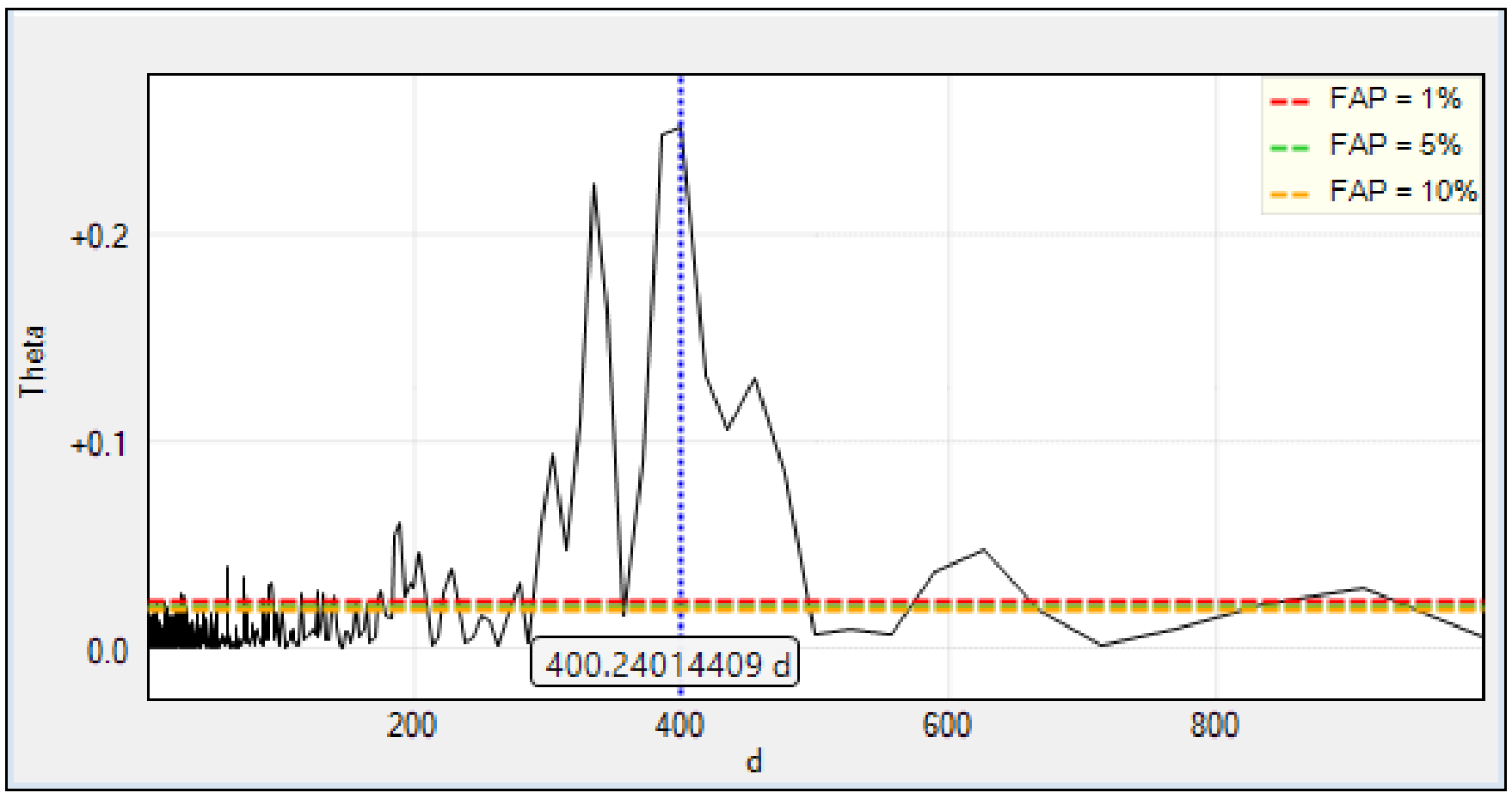

Figure 5: Peranso GLS Periodogram of the KWS-V data

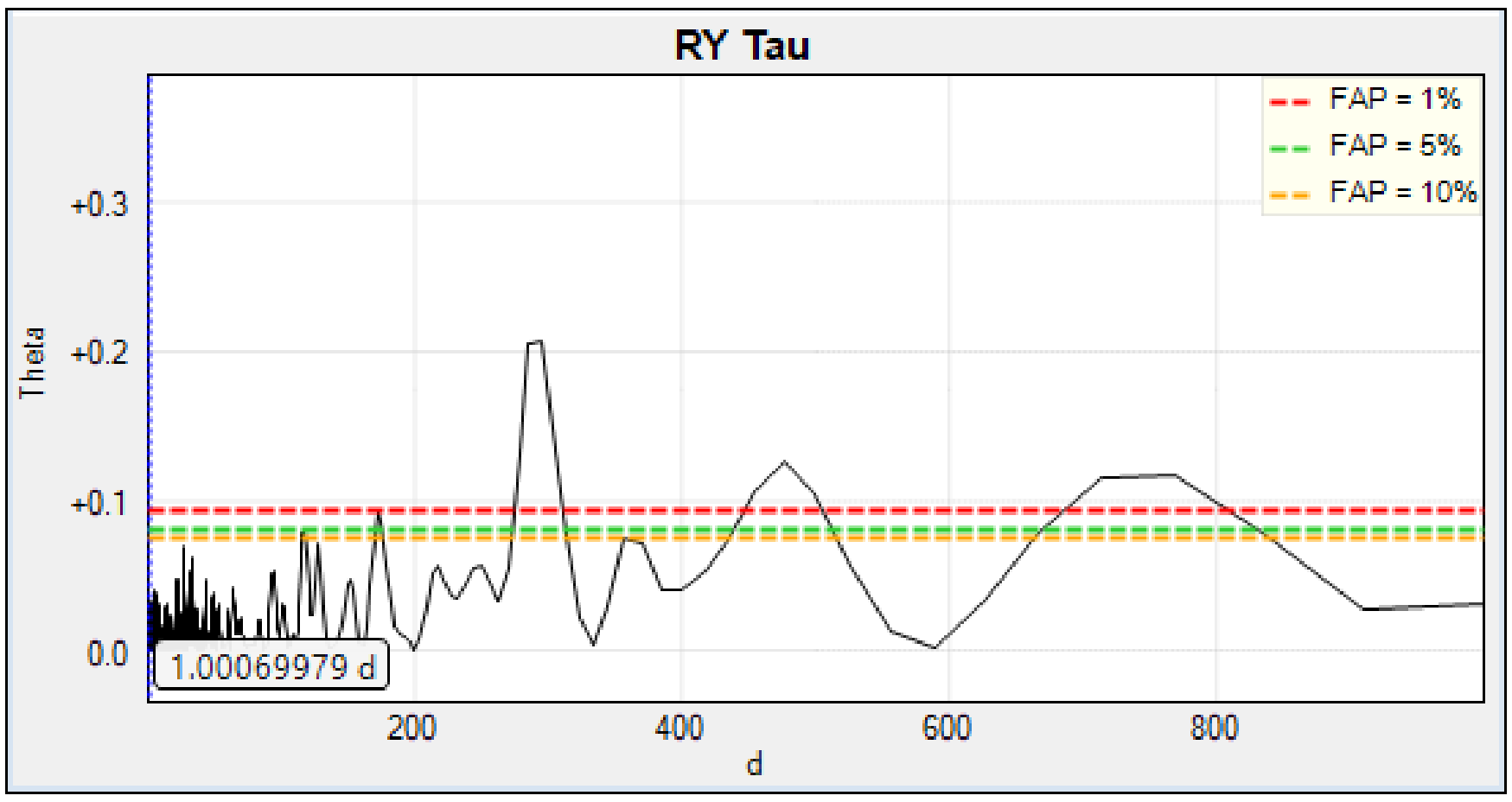

Figure 6: Peranso GLS Periodogram of the ASAS-SN V data

Since no clear periodicity is evident, no ephemeris calculation was performed in this case. Accordingly, the variability appears to be predominantly irregular, but with a sufficient amplitude of at least 1.5 mag to explain the optical variability of the surrounding reflection nebula.

**Short time variability of the nebula (ZTF frames)**

To further investigate the variability of the reflection nebula, we downloaded ZTF images via the IRSA server. We created an animation composed of approximately 450 individual frames from the Zwicky Transient Facility (ZTF), covering the period from 2018 to 2025. For this animation, the frames in the zr band were used, as the reflection nebula is significantly more prominent in this band than in zg. The time intervals between individual images are not uniform:
https://zenodo.org/records/18273567 , at half speed: https://zenodo.org/records/19117936
We encourage the reader to inspect the accompanying movie, where the reported variations are clearly visible and can be followed particularly well in a time-resolved manner. Obvious variations in the brightness and shape of the nebula are evident throughout the entire period. In the following, a particularly prominent episode of these changes between October and November 2018 is examined in detail.

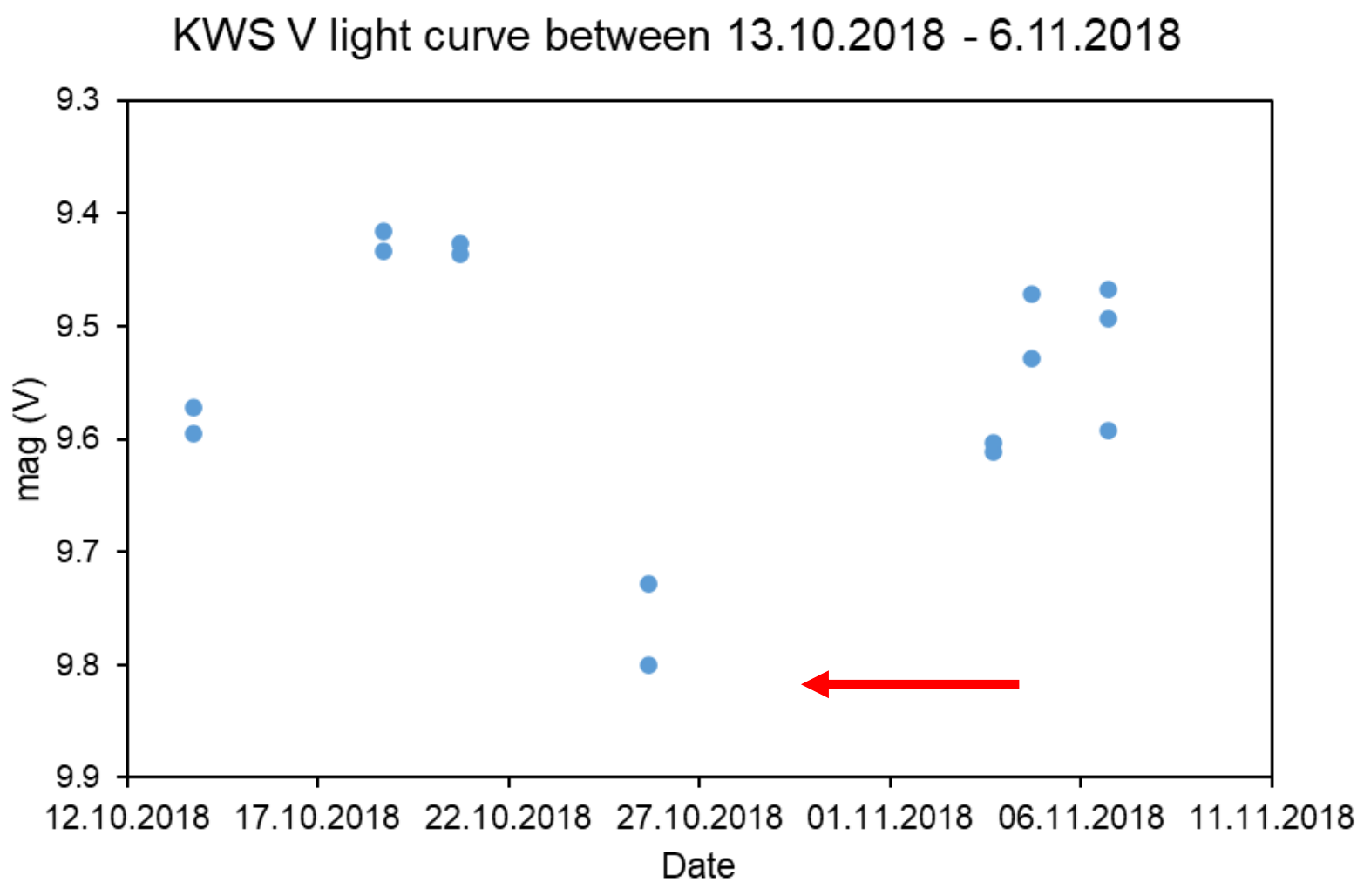

Figure 7: KWS V light curve of RY Tau between 13.10.2018 and 6.11.2018

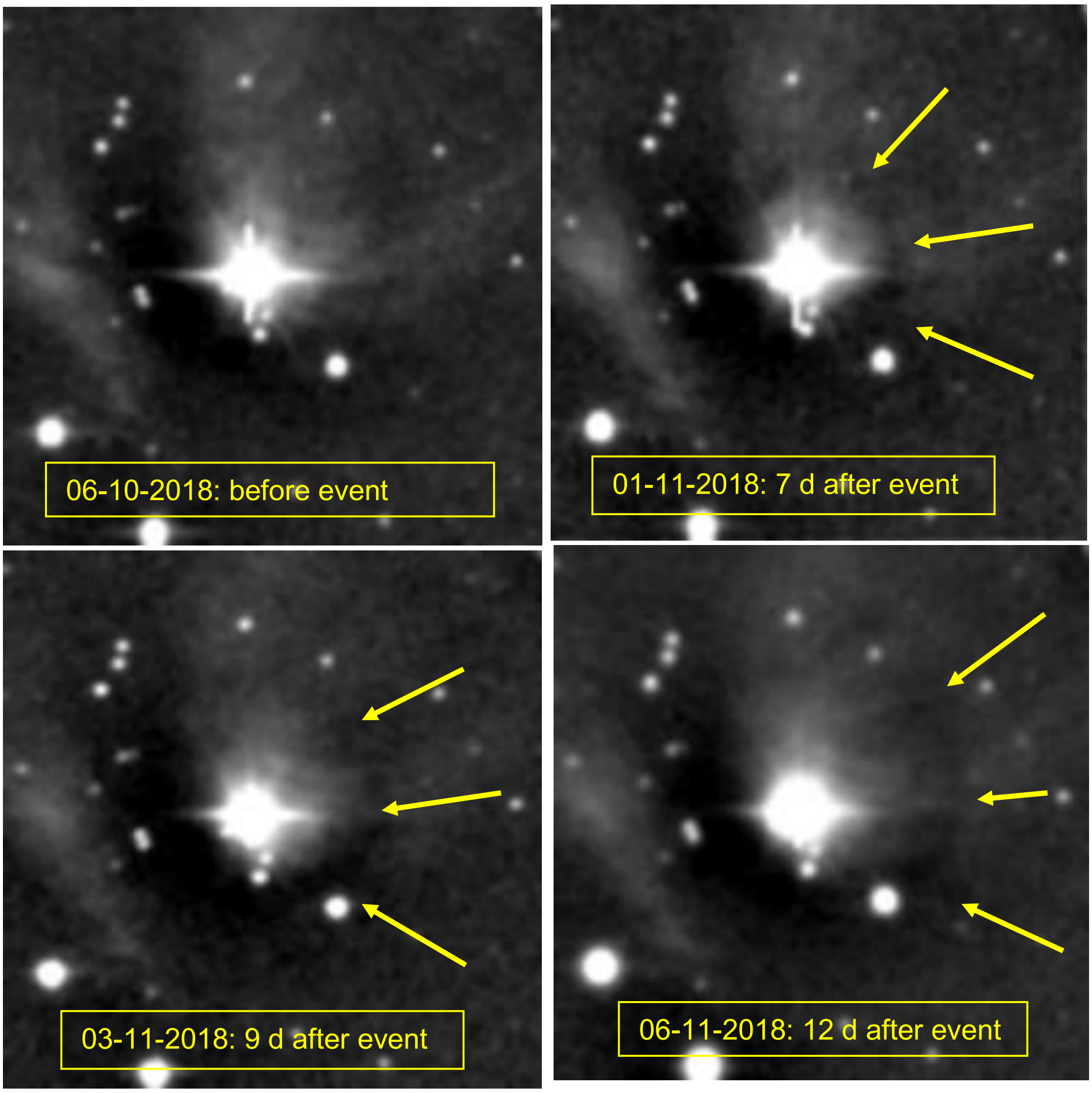

Figure 8 ZTF-r images before event (top left), 7 days after event (top right), 9 days after event (bottom left), 12 days after event (bottom right) - all images are contrast enhanced. The apparent displacement of the dimming front between 01 November 2018 and 06 November 2018 is approximately 23 arcsec, as indicated in the figure. All panels are shown with the same orientation; north is up and east is left.

The KWS light curve from the corresponding time interval shows that the brightness remained at around 9.5 mag, with a pronounced fading event around 25 October 2018 (Figure 7). This relatively short brightness decrease suggests the presence of a “dark light echo”, i.e., a decrease in scattered-light illumination propagating across the nebula, which can indeed be observed in the following weeks in the ZTF r-band images (Figure 8). The images obtained 7, 9, and 12 days after the event show an expanding darker front, marked by arrows. Since the front appears well defined, the velocity of the light echo can be estimated using the Gaia DR3 parallax as a distance proxy.

To estimate the propagation of the dimming front, we compared the ZTF r-band images from 1 November 2018 and 6 November 2018 and identified the same front segment on the western side of the nebula by its curvature and its position relative to adjacent nebular structures. The angular displacement was measured along the direction of maximum apparent motion in the astrometrically calibrated images, using the WCS-based plate scale, and is approximately 23 arcsec over 5 days. Owing to the diffuse morphology of the front, the uncertainty is dominated by the placement of the front boundary and is estimated to be about ±2 arcsec, corresponding to an apparent speed of roughly 3.4–3.8 c. This apparent superluminal motion is consistent with a light-echo effect, in which neither matter nor information travels faster than light. Instead, the effect arises from light-travel-time differences, as stellar radiation reaches different scattering surfaces at different times, causing the illuminated structures to appear to propagate faster than the speed of light across the nebula. A comparison of different regions of the nebula shows that the apparent propagation of the light echoes is not uniform: structures on the left evolve more slowly than those on the right and above. This indicates a three-dimensional geometry, with foreground scattering material on the right and above producing enhanced forward scattering and apparent superluminal motion, while the left-side structures are more laterally distributed with respect to the line of sight.

**Long term variability (Applause plates)**

For the investigation of the long-term brightness evolution of vdB 27, digitized photographic plates from the APPLAUSE[3] (Archives of Photographic Plates for Astronomical USE) project [17] were used. APPLAUSE provides high-resolution scans of historical astronomical plates from several observatories, including the Hamburg Observatory. The digitized material allows a direct comparison of archival observations spanning more than a century under homogeneous digital conditions.

The earliest plate shown in Figure 9 was obtained on 10 November 1923 with the 1 m reflector of the Hamburg Observatory (plate format 13 × 18 cm, Agfa Isorapid emulsion). The image obtained on 5 August 1981 was taken with the Hamburg Schmidt telescope (24 × 24 cm plates, Kodak 103a-E emulsion). These photographic data are compared with a modern CCD image from the ZTF g-band (2025). When visually comparing the three epochs, the main body of the nebula surrounding the variable star appears relatively brighter in the more recent images compared to the left-hand subregion indicated by the yellow arrow. However, this impression must be interpreted with caution. Differences in photographic emulsions, spectral sensitivity, plate characteristics, exposure times, and scanning procedures may influence the apparent brightness distribution. In addition, the short-term variability of both the central star and the nebular illumination is not known for the specific historical observing dates. Therefore, the observed contrast differences can only be regarded as suggestive of possible long-term structural or illumination changes, but they do not provide conclusive evidence for secular variability.

[3] https://www.plate-archive.org

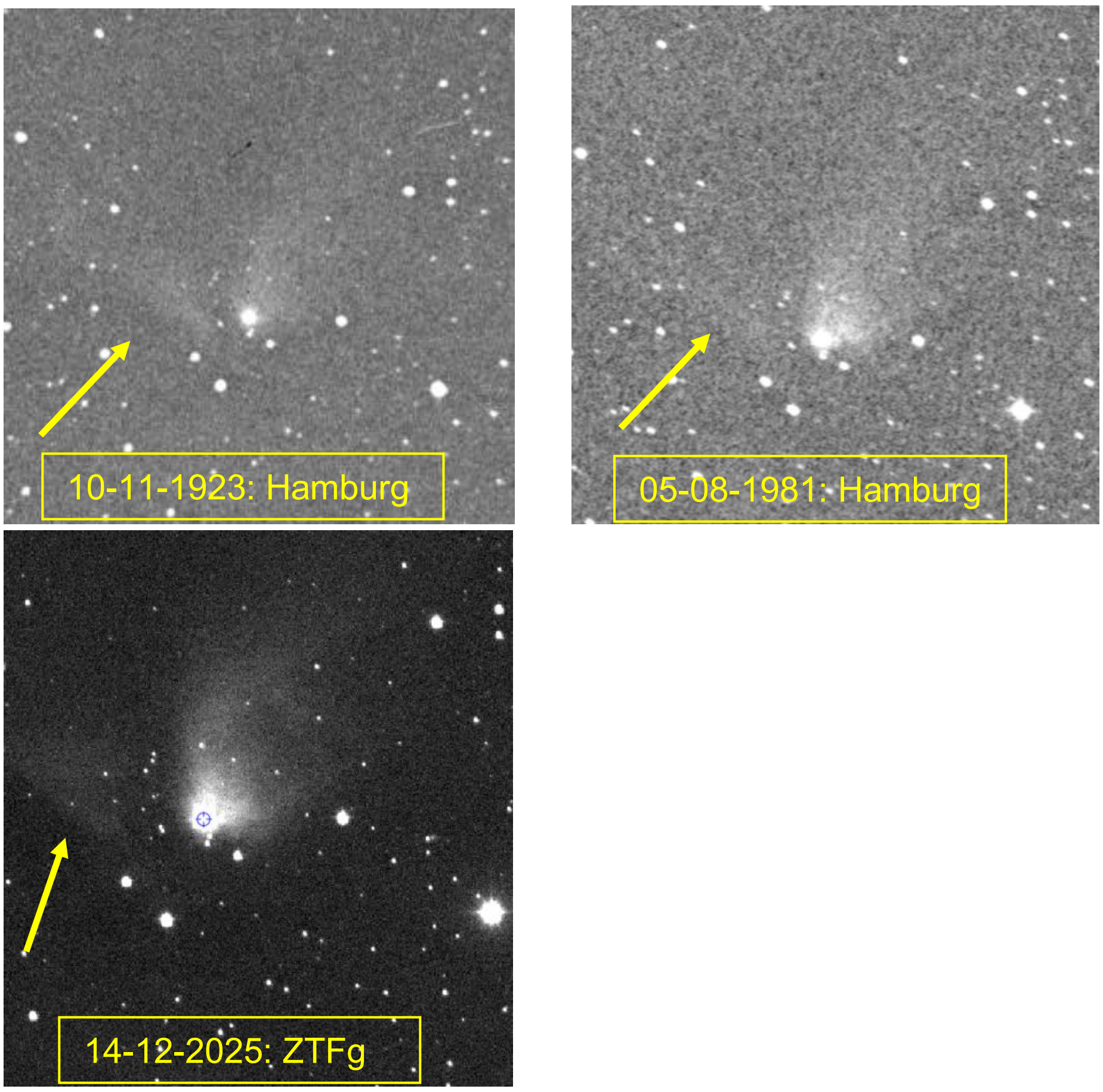

Figure 9: Comparison of images from 1923 (top left, Hamburg Observatory, 1 m reflecting telescope, digitized plate), 1981 (Hamburg Schmidt telescope, digitized plate), and 2025 (ZTF g-band). The yellow arrows mark the respective left-hand subregion discussed in the text.

## 4 Conclusion

RY Tau and its surrounding reflection nebula vdB 27 exhibit clearly detectable brightness variations in both the central star and the nebular structures, which are consistent with a light-echo effect. The characteristic timescales of these variations range from a few days to several weeks. Historical plates hint at possible secular changes, but the heterogeneity of bandpasses, emulsions, and image characteristics precludes a firm conclusion. This study therefore adds another object to the rare class of variable reflection nebulae. Further observations of such systems may provide valuable constraints on the geometry and scattering properties of circumstellar environments and reveal information about the direction-dependent variability of the illuminating star and the three-dimensional structure of the scattering dust.

## Acknowledgements

This research has utilized the SIMBAD/VIZIER database and Aladin, operated at CDS, Strasbourg, France, the International Variable Star Index (VSX) database, operated at AAVSO, Cambridge, Massachusetts, USA, the Zwicky Transient Facility (ZTF) and the SAO/NASA Astrophysics Data System, USA and the DSS Digitized Sky Survey (STScI). This research has made use of the NASA/IPAC Infrared Science Archive, operated by the Jet Propulsion Laboratory, California Institute of Technology, under contract with NASA. This research has also made use of the APPLAUSE database, funded by the German Research Foundation (DFG), the Leibniz Institute for Astrophysics Potsdam (AIP), Dr. Remeis Observatory Bamberg (University of Erlangen–Nürnberg), the Hamburg Observatory (University of Hamburg), and Tartu Observatory. The authors thank the referees for their helpful comments.